\documentclass[aps,prl,reprint,groupedaddress,showpacs]{revtex4-1}
\usepackage{graphicx}  
\usepackage{dcolumn}   
\usepackage{bm}        
\usepackage{amssymb}   
\usepackage{amsmath}
\usepackage{color}

\begin{document}


\title{Motional Averaging of Nuclear Resonance in a Field Gradient}


\author{Nanette N. Jarenwattananon}
\author{Louis-S. Bouchard}
\email[]{bouchard@chem.ucla.edu}
\affiliation{Department of Chemistry and Biochemistry, University of California Los Angeles, 607 Charles E Young Drive East, Los Angeles, CA 90095-1059}


\date{\today}

\begin{abstract}
The traditional view of nuclear-spin decoherence in a field gradient due to molecular self-diffusion is challenged on the basis of temperature dependence of the linewidth, which demonstrates different behaviors between liquids and gases. The conventional theory predicts that in a fluid, linewidth should increase with temperature; however, in gases we observed the opposite behavior. This surprising behavior can be explained using a more detailed theoretical description of the dephasing function that accounts for position autocorrelation effects.
\end{abstract}


\maketitle

\begin{figure}
\includegraphics[width=4cm]{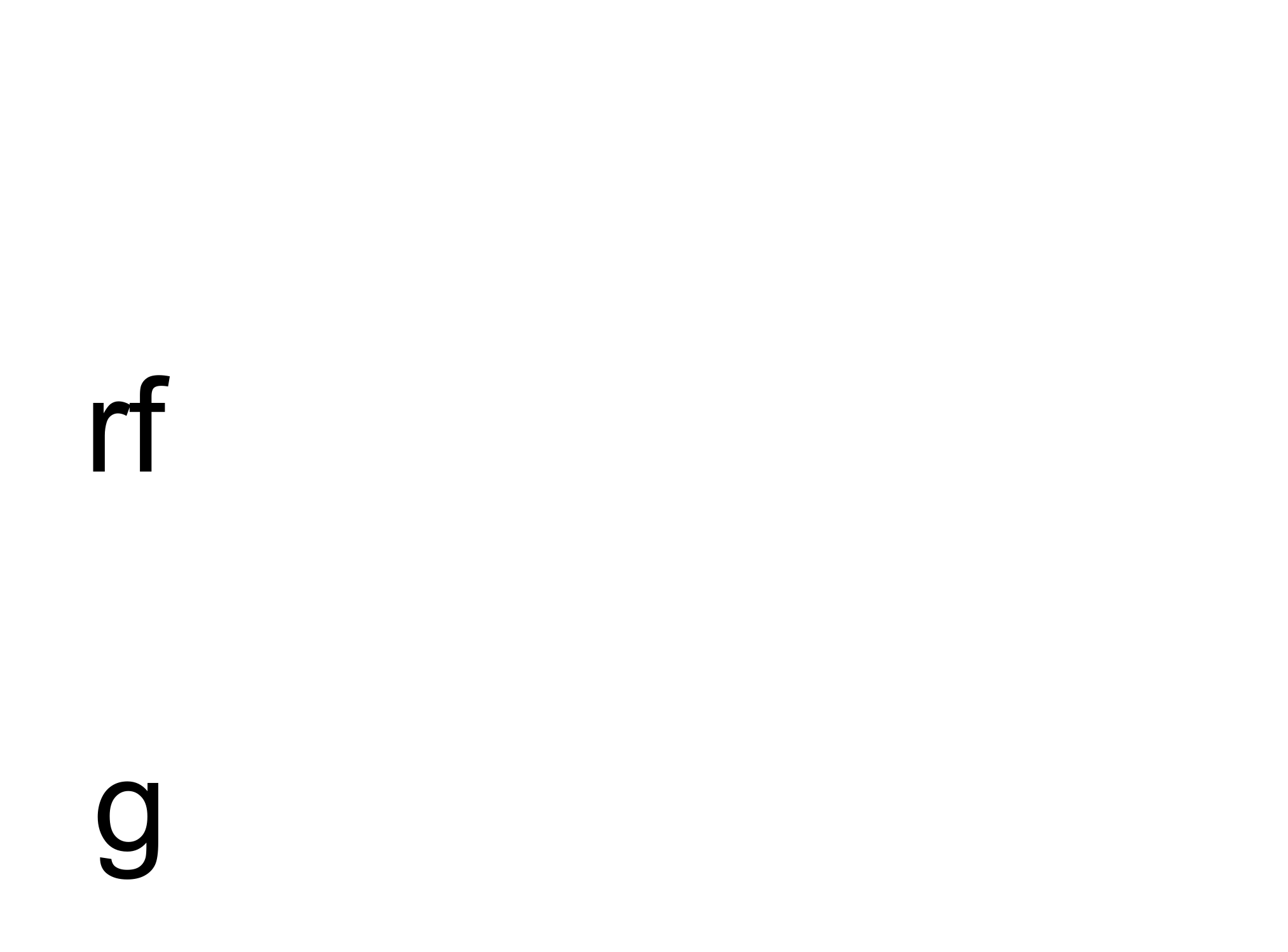}
\caption{\label{pulsesequence} Measurement of molecular self-diffusion in a constant field gradient. A 90$^{\circ}$ radiofrequency (rf) pulse tips the magnetization. This resulting nuclear induction signal is measured in a constant gradient of amplitude $g$. The time evolution of magnetization, as described by the textbook~\cite{stejskal1965, callaghan1991, callaghan2011} expression~(\ref{eq:1}), interrogates the molecular self-diffusion process.}
\end{figure}

For over six decades, diffusion-weighted nuclear magnetic resonance (NMR) has been the flagship experiment for measurements of molecular self-diffusion in free or confined geometries. Diffusion-based NMR experiments have a wide range of applications from porous media~\cite{bib:callaghannature1991}, catalysis~\cite{gladden1993}, materials science and chemistry~\cite{cohen2005} to biomedicine~\cite{lebihan2003}. Consider the simple experiment shown in Fig.~\ref{pulsesequence}, where the nuclear induction signal is read out in the presence of a magnetic-field gradient. The gradient modulates the magnetization spatially along the gradient's direction (assuming a sufficiently strong external field so that the gradient is unidirectional).  Time evolution of this magnetization in the presence of diffusion effects provides a direct and unambiguous measurement of the self-diffusion process.  In the traditional description of molecular self-diffusion~\cite{bloembergen1948, hahn1950, carr1954, torrey1956, kubo1954}, a molecule undergoes a random walk whereby at each time step, the nuclear spins accumulate phase increments that are randomly drawn from a normal distribution.  In the presence of a magnetic-field gradient, the decoherence of the nuclear induction signal $S(t)$ follows the well-known textbook expression~\cite{bloembergen1948, hahn1950,torrey1956, kubo1954,carr1954,callaghan1991}:
\begin{equation}
S(t) = \exp{(-(1/3)\gamma_n^2g^2Dt^3)}, 
\label{eq:1}
\end{equation}
where $\gamma_n$ is the nuclear gyromagnetic ratio, $D$ is the self-diffusion coefficient, $g$ is the applied gradient strength, and $t$ is time. The $t^3$ dependence has been extensively validated and has been utilized to measure molecular self-diffusion coefficients in a wide variety of liquids~\cite{stejskal1965, callaghan1991, callaghan2011}. In a gas, however, the situation is more complicated. The assumption of a normally-distributed phase accumulation at every time step is difficult to justify in light of the fact that gas molecules undergo much more rapid motion than in liquids, due to much longer free displacements between collisions. The farther the molecular displacements along the direction of the gradient, the faster the nuclear spins will lose memory of their immediate environment. This memory loss should be expected to enter the description of the decoherence process. Surprisingly, this simple aspect of free diffusion is still lacking a thorough experimental verification.

The NMR signal, $S(t)$, from an ensemble of spins initially located at $x(0)$ is given by the expectation value of the phase factor: 	
\begin{equation}
S(t)=\left\langle \exp \left( i \int_0^t \omega(t') dt' \right)  \right\rangle,
\end{equation}
where $\omega(t)$ describes the time-dependent resonance frequency offset (in the rotating frame).  In the presence of a magnetic field gradient $g$, the resonance frequency is $\omega(t)=\gamma_n g x(t)$, where $x(t)$, the position of the nuclear spin after time $t$, is a random process.  If we assume a Gaussian random process that is stationary in the wide sense, this expectation value takes the form~\cite{kubo1954}: 
\begin{equation} 
\exp \left( i \gamma_n g \int_0^t \langle x(t') \rangle dt' -  \gamma_n^2 g^2 \int_0^t \langle x(t') x(0) \rangle (t-t') dt' \right). 
\label{eq:2}
\end{equation}

\noindent The first term, $\exp( i \gamma_n g \int_0^t \langle x(t') \rangle dt')$, encodes the position in the phase of the spins~\cite{nishimura1996} after undergoing displacement but does not lead to decoherence unless the diffusion is not free.  The second term describes signal decay, and consequently, affects  linewidth.  More complicated random processes may lead to higher-order terms, but we shall limit our discussion to the case where $x(t)$ is a wide-sense stationary Gaussian random process.

Over the timescale $t$ of the NMR measurement, liquid phase molecules experience displacements that are much smaller than those in the gas phase. We may write this as $x(t) \approx x(0)$, for a liquid.  This approximation, known as the Einstein-Fick limit, indicates that the position autocorrelation function can be approximated by the mean-square displacement:
\begin{equation}
\langle x(t)x(0) \rangle \approx \langle x(t)x(t) \rangle = 2Dt,
\label{eq:3}
\end{equation}
where $D$ is the self-diffusion coefficient.  It is known from experiments that in liquids, this limit holds~\cite{stejskal1965, callaghan1991, callaghan2011}.  However, $D_{liquid}$ and $D_{gas}$ differ by three orders of magnitude, and it is unclear if this approximation also holds for gases.  Away from the Einstein-Fick limit, contributions from ballistic transport become significant.  We note that substitution of the limit~(\ref{eq:3}) into the second term of (\ref{eq:2}) recovers the result (\ref{eq:1}) as a special case.

In this Letter, we show that~(\ref{eq:1}) does not hold for gases based on the analysis of linewidth as function of temperature ($T$).  Because $D(T)$ increases with $T$,~(\ref{eq:1}) predicts that linewidth should increase with $T$.  We find that gases instead undergo line narrowing with temperature. It is unclear by cursory inspection of~(\ref{eq:1}) how gases may differ from liquids, especially in light of the fact that the dependence of line broadening on diffusion coefficient has been verified experimentally in several studies (see, for example,~\cite{pines1955, stejskal1965, stepisnik1993}). The key to establishing this distinction is a closer look at the temperature dependence of the line broadening mechanism, as explained below.

In the NMR experiment, nuclear spins are well isolated from the lattice and do not depolarize or randomize their phases when undergoing molecular collisions, in contrast to collisional broadening mechanisms in optics. Thus, the description of line broadening in such a ``weak collision" regime reflects the histories of molecular displacements.  The simplest way to account for this is through a position autocorrelation function. Suppose that the particle displacements are modeled using a generalized Langevin equation (GLE) with memory kernel:
\begin{equation}
 M \dot{v} + \int_0^t \Gamma(t-t') v(t') dt' = \eta_f(t), 
\label{eq:4} 
\end{equation}
where $M$ is the mass of the diffusing particle, $\Gamma(t)$ is a memory kernel, $v(t)=\dot{x}(t)$ is the particle velocity, $\dot{v}$ is its acceleration, and $\eta_f(t)$ is a stochastic force.  The GLE has been validated experimentally for Brownian particles ($M \gg m$, where $m$ is the mass of fluid particles); for example, in the studies~\cite{li2010,huang2011,franosch2011,kheifets2014} $M$ was $10^{10}$ times larger than $m$.  So while the GLE was not designed to model self-diffusion processes, it can be invoked to model viscous drag effects via the memory kernel.  In what follows, we shall set $M=10^{10} \cdot m$, which is the only regime we are aware of, where the GLE has been validated experimentally based on direct measurements of individual histories (namely, in Refs.~\cite{li2010,huang2011,franosch2011,kheifets2014}).

By the fluctuation-dissipation theorem, the stochastic force $\eta_f(t)$ describes colored noise, $\langle \eta_f(0) \eta_f(t) \rangle=k T \Gamma(t)$, where $k$ is Boltzmann's constant. The time-correlation function $\langle x( t) x(0) \rangle$ is obtained from $\langle v( t) v(0) \rangle$ by integrating twice the velocity autocorrelation function
\begin{equation}
\langle v( t) v(0) \rangle = - \frac{d^2}{dt^2} \langle x( t) x(0) \rangle.
\label{eq:5}
\end{equation}
\noindent Projecting equation~(\ref{eq:4}) with the operator $\langle v(0), \cdot \rangle$ yields the deterministic equation:
\begin{equation}
M \langle v(0) \dot{v}(t) \rangle + \int_0^t \Gamma(t-t')  \langle v(0) v(t') \rangle dt'=0.
\label{eq:6}
\end{equation}
\noindent Integrating this velocity autocorrelation function once
\begin{equation}
\nu(t) = \int_0^t  \langle v(0) v(t') \rangle dt',
\label{eq:7} 
\end{equation}
\noindent  and using the equipartition theorem, $\langle v(0) v(0) \rangle=k T/M$, as the initial condition, we get:
\begin{equation}
M \dot{\nu}(t)  + \int_0^t \Gamma(t-t') \nu(t') dt' =  k T. 
\label{eq:8}
\end{equation}
For the memory kernel to describe the delayed response of the surrounding fluid, we choose the Ornstein-Uhlenbeck process: 
$$\Gamma(t)=(\gamma^2/m) \exp (-\gamma t/m),$$ 
\noindent where $\gamma$ is the friction coefficient which is proportional to the viscosity of the medium and $m$ is the mass of molecules in the surrounding medium causing friction.   Denoting $\zeta_{-,+}=(\gamma/2m)(1 \mp \sqrt{1-4m/M})$, we obtain the solution to equation~(\ref{eq:8}):
\begin{widetext}
\begin{align}
\nu(t)  = \frac{ k T}{M} \Bigl\{  \frac{ \gamma}{ m \zeta_- \zeta_+ } +  \frac{1}{\zeta_+ - \zeta_-} \Bigl[ \Bigl(1-\frac{\gamma}{m\zeta_+} \Bigr)e^{ - \zeta_+ t}  - \Bigl(1-\frac{\gamma}{m\zeta_-} \Bigr)e^{- \zeta_- t} \Bigr] \Bigr\}.
\label{eq:9}
\end{align}
\end{widetext}

\noindent From this we get the position autocorrelation function,
\begin{widetext}
\begin{align}
\langle x(t) x(0) \rangle = \frac{kT}{M(\zeta_+ - \zeta_-)} \Bigl[\zeta_+^{-1} \Bigl(1-\frac{\gamma}{m\zeta_+} \Bigr) e^{ - \zeta_+ t}   - \zeta_-^{-1} \Bigl(1-\frac{\gamma}{m\zeta_-} \Bigr)  e^{-\zeta_- t} \Bigr]. \label{eq:10}
\end{align}
\end{widetext}
\noindent This result was also derived by N{\o}rrelykke \cite{norrelykke2011} using a different method. In the Einstein-Fick approximation, $x(0)\approx x(t)$, and this position autocorrelation function reduces to $2Dt$.  Equation (\ref{eq:10}) generalizes $S(t)$ outside the Einstein-Fick limit.  There are three distinct regimes: overdamped ($M>4m$), critically damped ($4m=M$), and underdamped ($M<4m$). Standard Brownian motion of large particles is strongly overdamped ($M\gg m$). The Einstein-Fick limit occurs when the ratio $\gamma t/m$ is sufficiently large to cause appreciable decay of the Ornstein-Uhlenbeck kernel.  Using $\gamma$ based on the Stokes' law and $t=40~\mu$s as the sampling time of the nuclear induction signal, we find that typical values of this ratio for liquids are $\gamma t/m \sim 1$ whereas for gases we have $\gamma t/m \ll 1$.

In the case of a gas, we may obtain the overall temperature by modeling $\gamma$ using the Stokes' law, $\gamma = 3\pi \eta_v d$ ($\eta_v$, shear viscosity of the medium; $d$, sphere diameter), which holds in the limit of low Reynolds numbers and invoking the Sutherland's formula~\cite{chapman1970} for $\eta_v$:
\begin{equation}
 \eta_v = \frac{ \mu_0 (T_0+C) (T/T_0)^{3/2}}{ T+C} \sim \frac{ T^{3/2}}{ T+C}, 
\label{eq:11}
\end{equation}
\noindent where $C$ is Sutherland's constant for the gas, $\mu_0$ is the viscosity at temperature $T_0$.  At low temperatures, $\eta_v \sim T^{3/2}$, whereas at high temperatures, $\eta_v \sim \sqrt{T}$.  By substituting this into~(\ref{eq:11}), we obtain at an expression for the envelope function of the signal decay for a gas which does not rely on the Einstein-Fick limit:
\begin{equation}
S(t) = \exp \left({-\gamma_n^2g^2 \kappa t} \right) 
\label{eq:12}
\end{equation}
\noindent with 
\begin{equation}
\kappa(T)= \frac{kT(-m \zeta_-^2 \zeta_+ - m \zeta_- \zeta_+^2 + \zeta_- \gamma + \zeta_- \zeta_+ \gamma + \zeta_+^2 \gamma)}{mM \zeta_-^3 \zeta_+^3}.
\end{equation}
\noindent The linewidth $\Delta f$ follows the power law:
\begin{equation}
  \Delta f \sim \begin{cases}
T^{-7/2}, & T<C \\
T^{-1/2}, & T> C
\end{cases},
\label{eq:13}
\end{equation}
which predicts a temperature dependence that is opposite (i.e., line narrowing with increasing temperature) to that based on self-diffusion in the Einstein-Fick limit (\ref{eq:1}).  An analogous expression in the case of liquids can be derived using a suitable model for the temperature dependence of the viscosity in a liquid.  However, this will not be needed here, because we shall see that the linewidth is essentially independent of temperature.
 
\begin{figure}[ht]
\includegraphics[width=8.0cm]{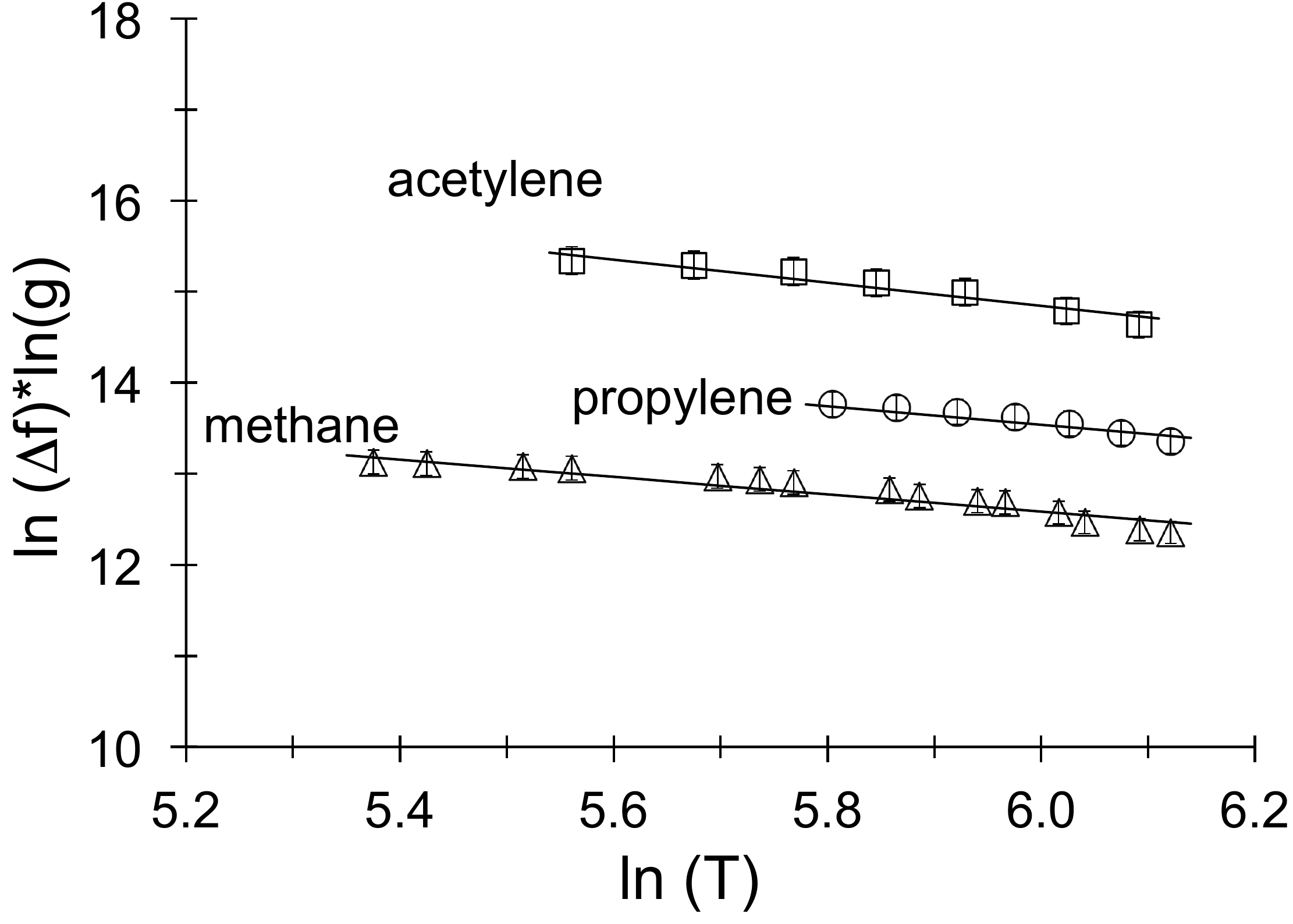}
\caption{\label{gases_lw_v_t}Validation of the $T^{-1/2}$ law in the high-temperature regime ($T>C$) for gases.  The values of $\ln T$ shown correspond to the temperature range $T=180-490$ K.   Three different gases were investigated: methane, acetylene, and propylene.  The temperature dependence on linewidth (scaled to gradient strength) was found to be $\Delta f \propto T^{-0.47 \pm 0.04}$ (averaged over the three gases).  Although different values of the applied gradient $g$ are shown here to avoid overlapping of the curves (methane, $g$=0.15~G/cm; acetylene, $g$=0.07~G/cm; propylene, $g$=0.1~G/cm), the scaled linewidth is independent of $g$.}
\end{figure}

\begin{figure}[ht]
\includegraphics[width=8.0cm]{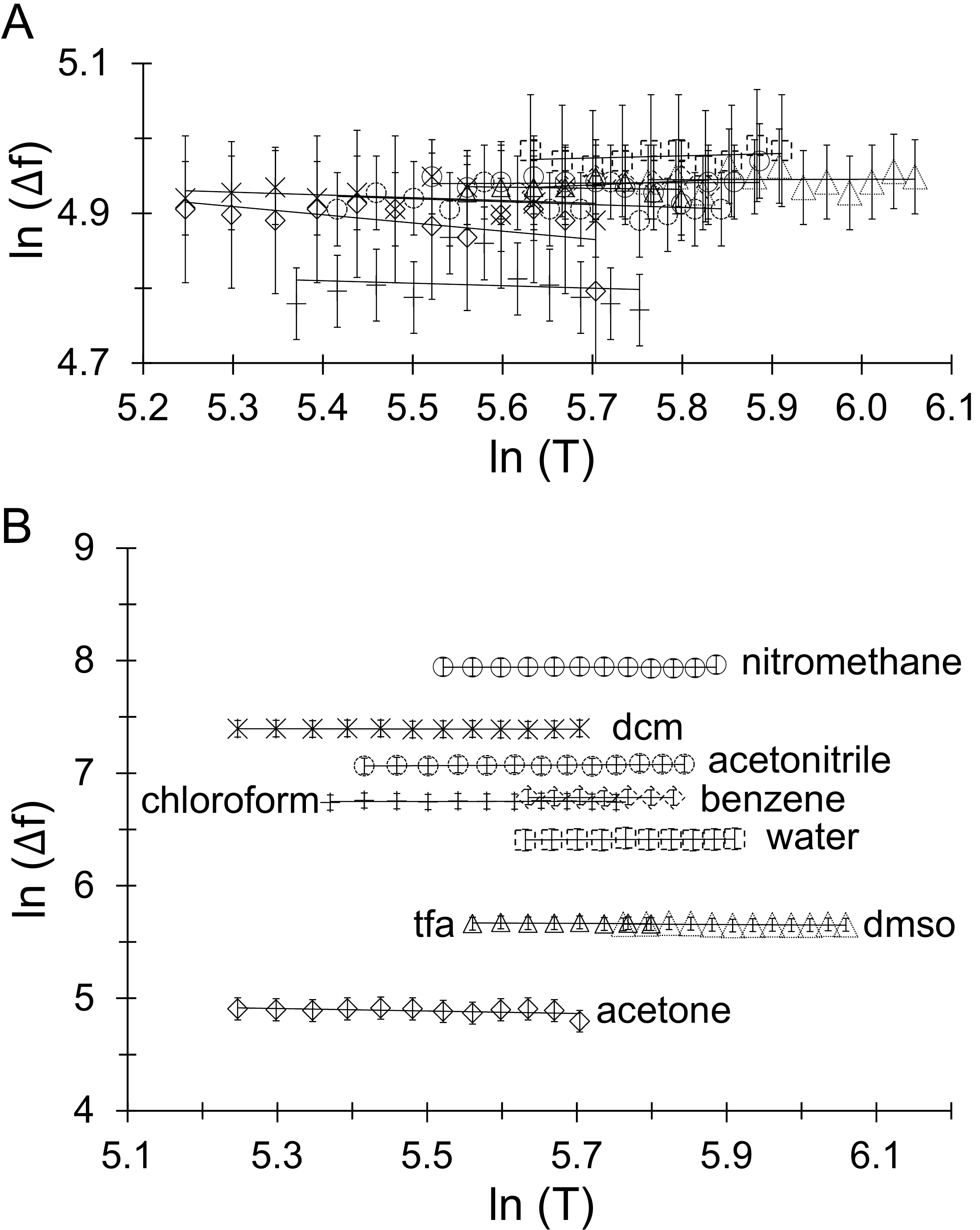}
\caption{\label{liquids_LW_v_T} Temperature dependence of linewidth ($\Delta f$) in liquids.  Nine different liquids were investigated, as shown by the different symbols.  The values of $\ln(T)$ shown span the temperature range $T=180-450$ K. A)  For a fixed gradient strength of $g$=0.05~G/cm, all linewidths were broadened by a similar amount, hence the overlap in the data. At fixed $g$, the linewidth did not exhibit any detectable dependence on temperature.  B) Increasing applied gradient strength did not alter the independence of linewidth on temperature. Applied gradient strengths were: nitromethane (1~G/cm), dichloromethane (dcm, 0.5~G/cm), acetonitrile (0.4~G/cm), chloroform (0.3~G/cm), benzene (0.3~G/cm), water (0.2~G/cm), trifluoroacetic acid (tfa, 0.1~G/cm), dimethyl sulfoxide (dmso, 0.1~G/cm), acetone (0.05~G/cm).   Symbol in A refers to the same liquid as in B. }
\end{figure}

Measurements of $\Delta f$ were carried out as a function of $T$ for three gases in the high temperature regime ($T>C$, as determined by the Sutherland's constant for each gas~\cite{supplementalmaterial,ammann1982,CRChandbook,sutherland1909}). The results are shown in Figure~\ref{gases_lw_v_t}.   The average exponent was found to be $-0.47\pm0.04$, in agreement with the theoretically predicted value of $-1/2$ in Eq. (\ref{eq:13}).  The low temperature regime ($T<C$) could not be investigated due to experimental limitations of our instrument.   A temperature dependence of linewidth could not be detected within experimental error for liquids, as shown in Figure \ref{liquids_LW_v_T}, where we investigated nine different liquids over the range 180-450 K. We note that the Sutherland's formula (and therefore, Equation~\ref{eq:12}) is applicable to gases only, so a lower limit on temperature for the liquids is imposed by the freezing points.

\begin{figure}
\includegraphics[width=8.0cm]{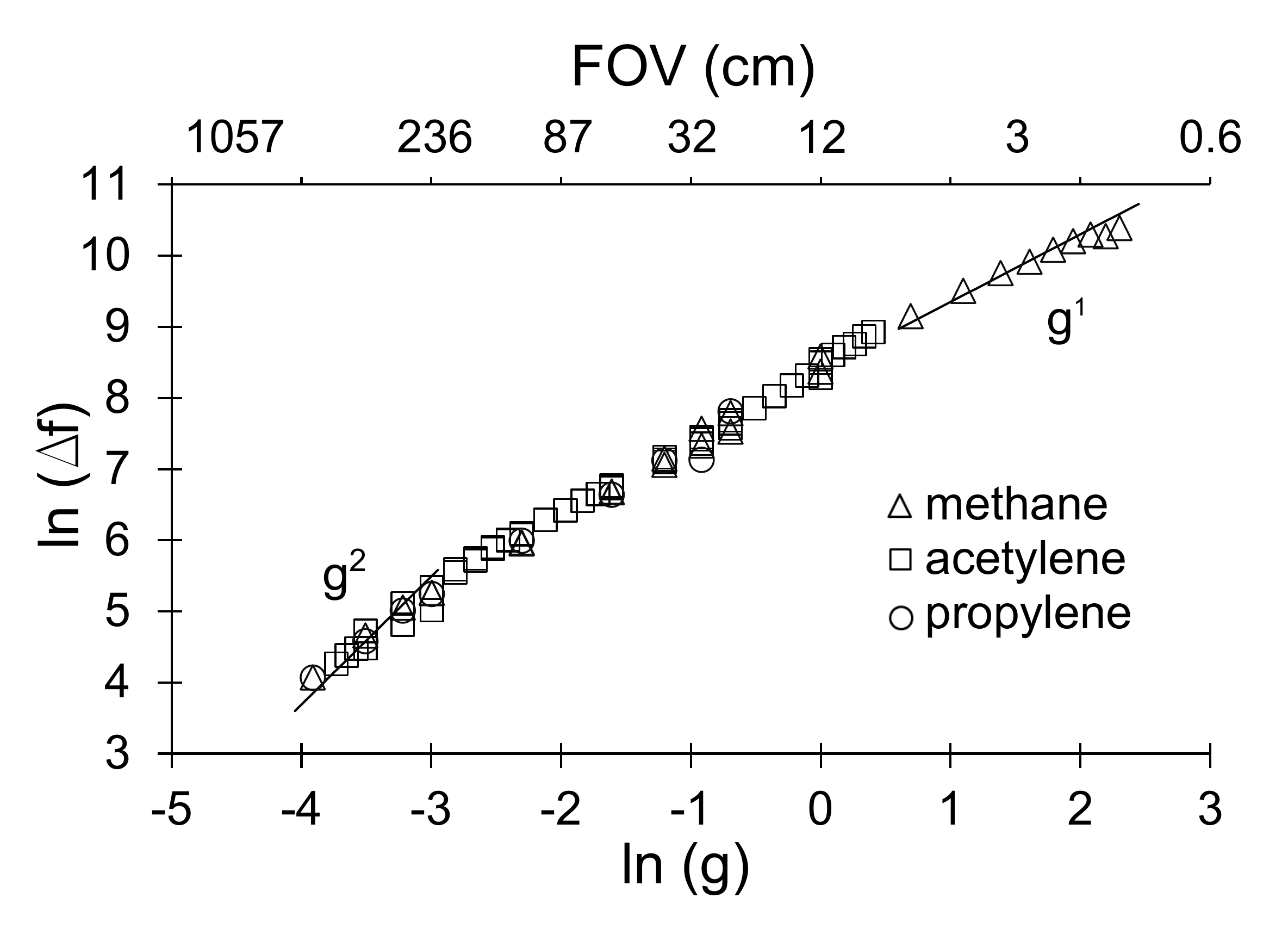}
\caption{\label{gases_LW_v_g} Dependence of linewidth ($\Delta f$) on applied gradient strength ($g$) for gases.   Three different gases were investigated.  Two different regimes are found: in the limit of strong applied gradients, $\Delta f$ scales as $g^{1.0 \pm 0.1}$ whereas for weak gradients $\Delta f$ scales as $g^{1.8 \pm 0.2}$. All data was acquired at ambient temperature.}
\end{figure}

\begin{figure}
\includegraphics[width=8.0cm]{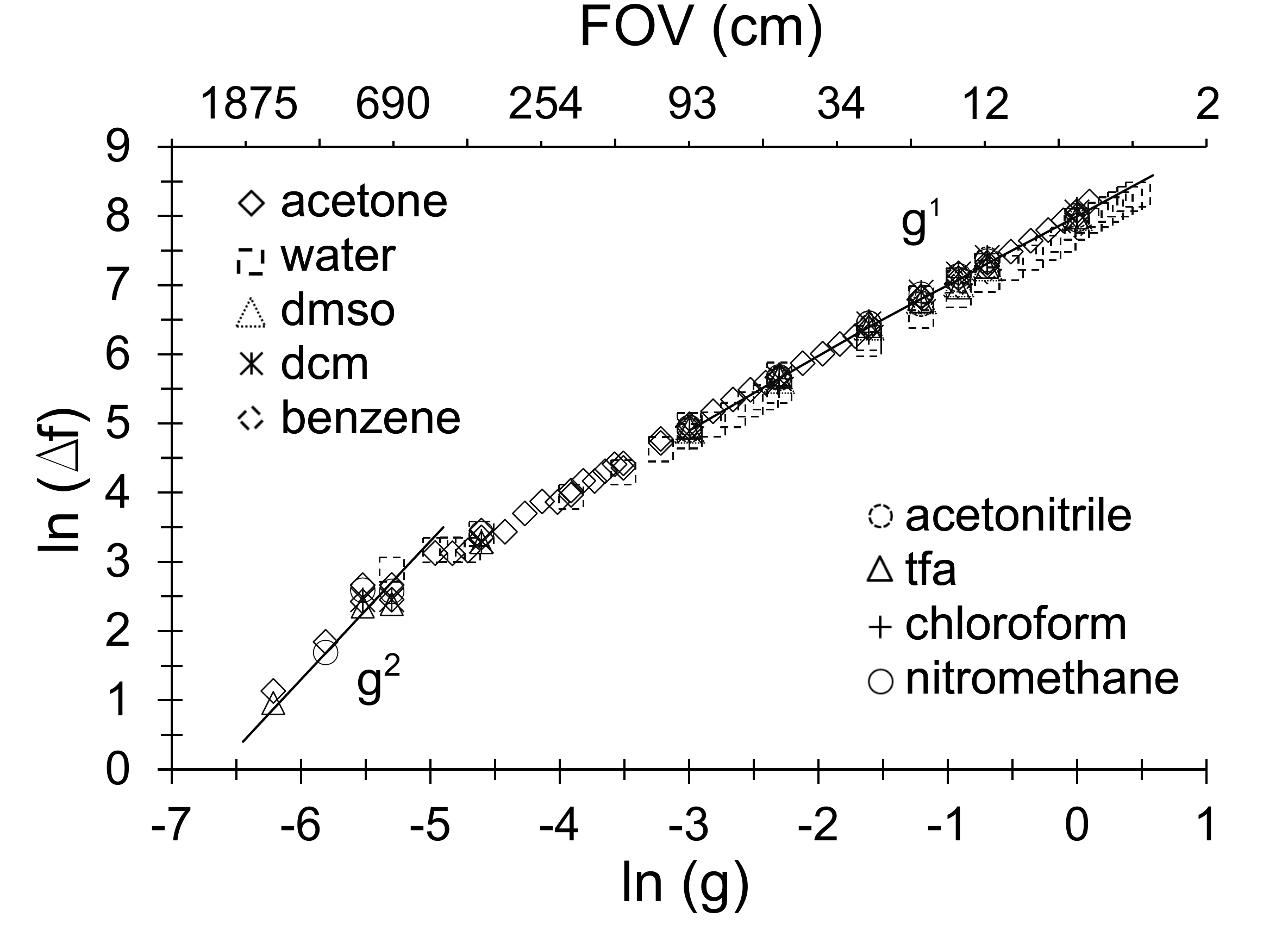}
\caption{\label{liquids_LW_v_g} Dependence of  linewidth ($\Delta f$) on gradient strength ($g$) for liquids.  Nine different liquids were investigated.  Two different regimes are found: in the limit of strong applied gradients, $\Delta f$ scales as $g^{1.1 \pm 0.1}$ whereas for weak gradients $\Delta f$ scales as $g^{2.0 \pm 0.2}$. All data was acquired at ambient temperature.}
\end{figure}

We now turn out attention to the gradient dependence of the line broadening, which, according to~(\ref{eq:12}), should be proportional to $g^2$. In experiments, however, we found two regimes:  in the limit of weak gradients, $\Delta f \propto g^2$, whereas for strong gradients, $\Delta f \propto g^1$ (see Figs.~\ref{gases_LW_v_g} and \ref{liquids_LW_v_g}). The $g^2$ gradient dependence is predicted according to (\ref{eq:12}); however, the $\Delta f \propto g^1$ is not.  Moreover, these two regimes apply to both liquids and gases, according to the results of Figs.~\ref{gases_LW_v_g} and \ref{liquids_LW_v_g}.   The emergence of the $g^1$ regime is likely due to the convolution of the line shape with the sample shape that arises under an applied gradient and forms the basis for frequency encoding in magnetic resonance imaging.  The full signal equation integrated over the sample shape is:
\begin{equation}
S(t) = \int  \exp(i \gamma_n g x t) \exp(-\alpha g^2 t) \rho (x) dx, \label{eq:14}
\end{equation}
\noindent where $\rho(x)$ describes the spin density profile along the $x$ direction (sample shape), $\exp(i \gamma_n g x t)$ is the frequency encoding and $\exp(-\alpha g^2 t)$ is the line broadening according to (\ref{eq:12}).  For the particular case where the sample shape $\rho(x)$ is Gaussian (a reasonable approximation for the sensitivity profile of a saddle coil, such as the one used in these experiments), the Fourier transform of (\ref{eq:14}) with respect to time is a convolution:
\begin{equation}
\tilde{S}(\omega) = \mbox{Gauss} \otimes  \mbox{Lorentz} = \mbox{Voigt},
\label{eq:15}
\end{equation}
\noindent in which the Gaussian is $\sim\exp(-x^2/2\sigma^2)$, and the Lorentzian is $\sim\Gamma^2/(x^2 + \Gamma ^2)$.  In terms of the full width at half maximum of the Gaussian ($f_G = 2\sigma \gamma g \sqrt{2 \ln{2}}$) and Lorentzian ($f_L = 2\Gamma = 2 \alpha g^2$) profiles, the width of the Voigt profile can be expressed as $f_v \approx 0.5 f_L + \sqrt{0.2 f_L^2 + f_G^2}$. 
Thus, in the frequency encoding regime, where sample shape effects dominate, the line broadening behaves as $g^1$ regardless of whether~(\ref{eq:1}) or (\ref{eq:12}) are used to describe diffusion effects. The frequency encoding regime is reached when the field of view FOV=$f_s/\gamma_n g$ ($f_s$, sampling rate; $g$, gradient amplitude), becomes comparable to the size of the rf-sensitive region ($\sim$ 1 cm in our experiments). Depending on the applied gradient strength, the experimental FOV ranges from $0.6$ to $1460$~cm. The FOV values corresponding to applied gradient ($g$) are indicated in the upper horizontal axes of Figs.~\ref{gases_LW_v_g} and \ref{liquids_LW_v_g}, where the two regimes, $g^1$ and $g^2$, are indicated. 

We have presented a revised expression for line broadening (\ref{eq:12}) that not only takes into account the autocorrelation effects in the diffusion process, but also suggests that self-diffusion processes in the NMR experiment may be described using a stochastic GLE; at least, as far as its temperature dependence is concerned.  The GLE~(\ref{eq:12}) enables a convenient description of the memory effects arising from the viscous drag effects, which were essentially missing from the traditional description (\ref{eq:1}). Such drag effects yielded the correct temperature dependence for gases and have been used in a recent publication to non-invasively map temperatures of gases during catalytic reactions~\cite{jarenwattananon2013}. The method could also be useful in the validation of heat-transfer models for gas-phase thermal exchange systems, which currently rely on numerical results from computational fluid dynamics models. Finally, we note that since the decay function~(\ref{eq:12}) involves the first power of time instead of its third power (\ref{eq:1}), (\ref{eq:12}) could have implications for the design of dynamic decoupling schemes for coherent quantum control~\cite{uhrig2007}.

See Supplemental Material~\cite{supplementalmaterial} at [URL to be inserted by publisher] for experimental methods, data analysis details, and sample data.

\begin{acknowledgments}
This work was partially funded by a Beckman Young Investigator Award and the National Science Foundation through grant CHE-1153159.
\end{acknowledgments}

\section{Supplementary Material (SM)}

\section{SM: Experimental Set-Up}
\subsection{Sample Preparation} Methane gas ($> 99\%$ purity) was purchased from Airgas, Inc. and used as provided. Acetylene($> 98\%$ purity) and propylene gases ($> 95\%$ purity) were purchased from Praxair, Inc. and used as provided. For the gas-phase experiments, a sealable J. Young NMR tube was evacuated to remove excess air, and filled with pure gas to 15 PSIA. Liquids were purchased from Sigma-Aldrich. For liquid-phase experiments, samples were degassed and flame-sealed in NMR tubes.

\subsection{Nuclear Magnetic Resonance Methods} Measurements were performed on a 14.1~T vertical bore Bruker AV 600 MHz NMR spectrometer equipped with a 5 mm broadband probe with a z-direction gradient. The sample was placed in the center of the NMR magnet, and the pulse sequence~(Fig. 1, main text) was applied. The receiver was placed in DQD mode (forward Fourier transform, quadrature detection of complex data). During acquisition, a linewidth-broadening magnetic field gradient of amplitude $g$ was applied.

\subsection{Temperature Control} Temperature is altered by the variable temperature (VT) unit, which controls the temperature of the sample by heating or cooling the surrounding air. Temperatures above ambient temperatures ($298$ K $<$ T $< 460$ K) are achieved by the probe's internal heater. Temperatures below ambient temperatures ($200 $K $< $T $< 298$ K) are achieved with a heat-exchange coil, by pre-cooling the gas in a liquid nitrogen bath. Due to experimental limitations in the VT-system, we were unable to achieve temperatures lower than 200 K or above 460 K. In order to determine the real sample temperature, each VT-temperature value was calibrated against a neat methanol standard for temperatures $200$~K $~<$~T~$<~300$~K (Eq.~\ref{eq1}) and a neat ethylene glycol standard for temperatures $301$~K~$<$~T~$<~460$~K (Eq.~\ref{eq2}):
\begin{equation}
T [K] = 409.0 - 36.54 (\Delta d) - 21.85 (\Delta d)^2
\label{eq1} 
\end{equation}
and
\begin{equation}
T [K] = 466.5 - 102.00 (\Delta d),
\label{eq2} 
\end{equation}
in which $(\Delta d)$ is the chemical shift difference between the two peaks of neat methanol or two peaks of neat ethylene glycol~[23]. We note that at colder temperatures (< 290K), temperatures tended to fluctuate more than at ambient and above ambient conditions.

\section{SM: Data Analysis and Sample Data} Fast Fourier transformation (FFT) and subsequent analysis with the MATLAB Curve Fitting Toolbox was performed in MATLAB R2010b (The MathWorks, Inc.; Natick, MA).

\subsection{Linewidth vs. temperature data} 
For gases, linewidth vs. temperature data for each gas was fit to a linear regression.  Only the high-temperature regime (i.e. above the Sutherland's constant C) could be studied due to limitations in the VT control. In Ref.~[21] the Sutherland's constant was calculated for each gas according to its viscosity~[21,24] and Sutherland's model [21,25]: 
	\begin{equation}
	\mu = \mu^\prime \left(\frac{T}{T^\prime}\right)^{3/2} \left(\frac{T^\prime + C}{T + C}\right), \label{eq3}
	\end{equation}
in which $\mu$ is the viscosity of a gas at temperature $T$, $\mu^\prime$ is the viscosity of that same gas at another temperature $T^\prime$, and $C$ is the Sutherland's constant for that gas. The values of $S$ were calculated to be 198 K, 237 K, and 292 K for methane, acetylene, and propylene respectively. For liquids, the lower bound of temperature was determined by the freezing point of the substance; all data above this temperature was fit to a linear regression. 

\begin{figure*}[h!]
\includegraphics[width=15cm]{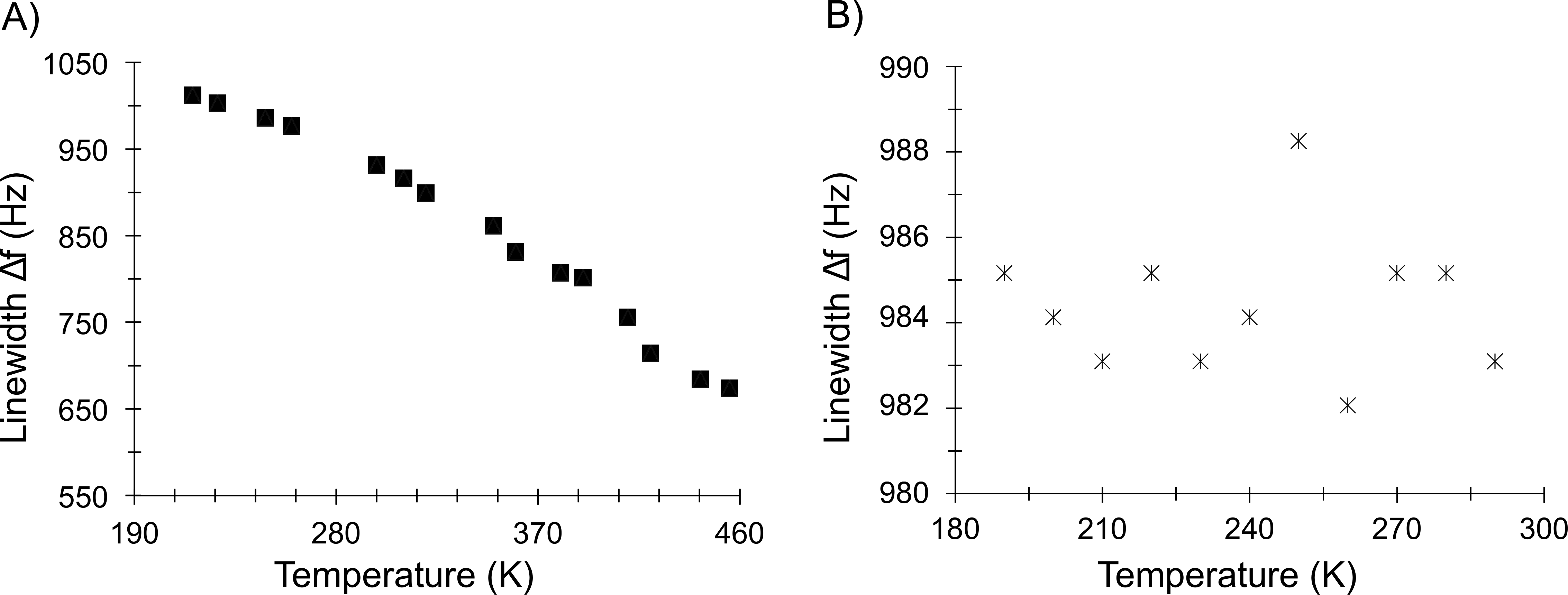}
\caption{\label{fig_s1} A) Raw linewidth vs. temperature data for methane gas. The applied gradient strength was $g = 0.15$ G/cm, with FOV = 78 cm. B) Raw linewidth vs. temperature data for liquid dichloromethane. The applied gradient strength was $g = 0.3$ G/cm, with FOV = 94 cm. }
\end{figure*}

Figure \ref{fig_s1}(A) is the raw data from a methane linewidth vs. temperature experiment. The gradient strength is $g = 0.15$ G/cm, yielding a FOV of 78 cm.  Figure~ \ref{fig_s1}(B) is the raw data from a dichloromethane linewidth vs. temperature experiment. The applied gradient strength was $g = 0.3$ G/cm, yielding a FOV of 94 cm. Linewidth for the gas-phase data follows a monotonic decrease as a function of temperature, whereas linewidth for liquid-phase data does not follow a clear trend.

\subsection{Linewidth vs. gradient data} For linewidth vs. gradient data, the two regimes were determined based on  the field of view (FOV), as calculated according to the following equation: FOV=$f_s/\gamma_n g$ ($f_s$, sampling rate; $g$, gradient amplitude). When the FOV becomes comparable to the size of the rf-sensitive region ($\sim$ 1 cm in our experiments), we are in the frequency-encoding ($g^1$) regime; when FOV $\gg$ 1 cm, we are in the non-frequency encoding, $g^2$ regime. The coefficient of the $g^1$ and $g^2$ slopes were determined by fitting data in the frequency-encoding regime and non-frequency encoding regimes respectively to a weighted quadratic function.

\begin{figure*}[h!]
\includegraphics[width=15cm]{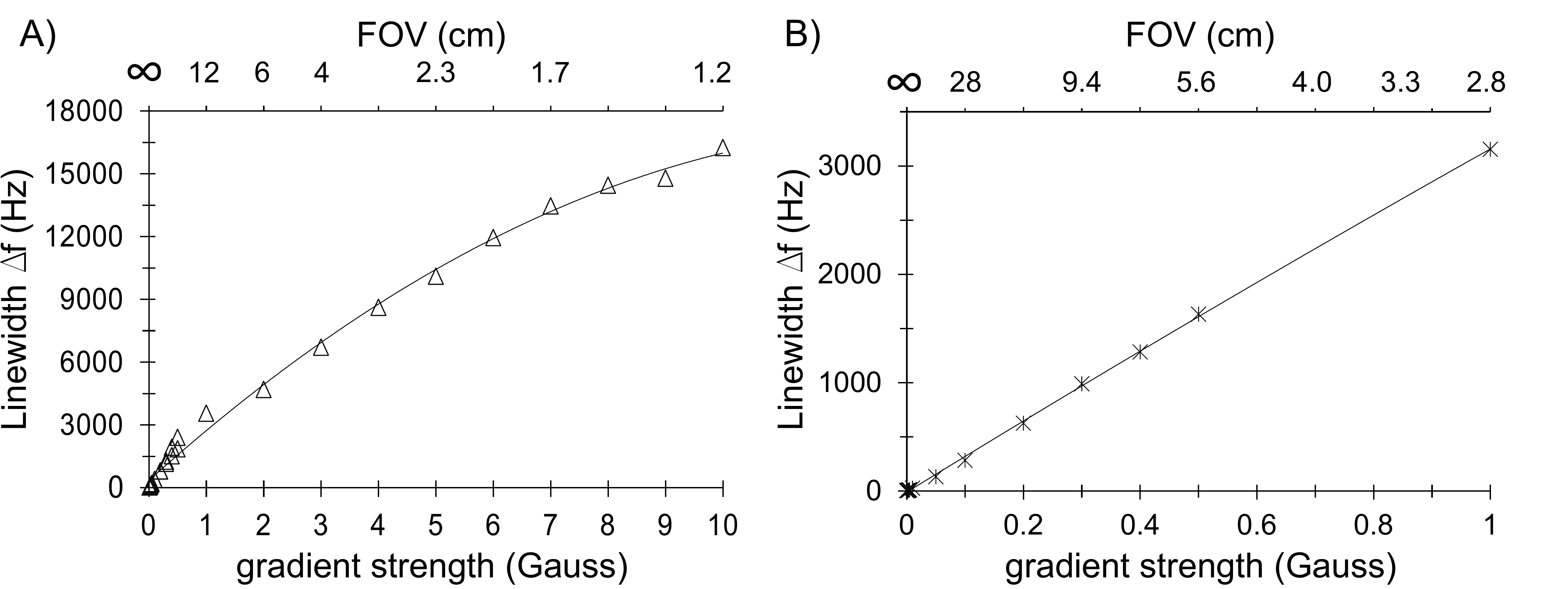}
\caption{\label{fig_s2} A) Raw linewidth vs. gradient data for methane gas, acquired at 220 K. B) Raw linewidth vs. gradient data for liquid dichloromethane, acquired at 300 K. The continuous curves drawn through the data points are guides to the eye, and are not actual fits.}
\end{figure*}

Figure~ \ref{fig_s2}(A) is the raw data from a methane linewidth vs. gradient experiment. The data was acquired at T = 220 K. Figure~ \ref{fig_s2}(B) is the raw data from a dichloromethane linewidth vs. gradient experiment. The data was acquired at T = 300 K. Although the quadratic curves are drawn as a guide to the eye, the data illustrate that linewidth monotonically increases as a function of gradient strength.

\subsection{Determination of Error Bars} At each temperature point, five scans were separately acquired and averaged to reduce the effects of temperature fluctuations over the time course of the experiment. For each substance studied, multiple experiments were performed on different days to eliminate the random fluctuations in field homogeneity day to day resulting from magnetic field drift.

\section{SM: References}
\noindent 21. S. Chapman and T.G. Cowling. The Mathematical Theory of Non-Uniform Gases: An account of the kinetic theory of viscosity, thermal conduction and diffusion in gases.  Cambridge University Press, \textbf{1939}. \\
23. C. Ammann, P. Meier, and A.E. Merbach. \textit{J. Magn. Reson.} \textbf{1982}, \textit{46}, 319-321. \\
24. W.M. Haynes, Ed. CRC Handbook of Chemistry and Physics, 95th Edition (2014-2015).  \underline{http://www.hbcpnetbase.com}\\
25.  W. Sutherland. \textit{Phil. Mag.} \textbf{1909}, \textit{17}, 320-321. \\

\end{document}